# The Analytical Expressions for a Finite-Size 2D Ising Model

**M.Yu. Malsagov, I.M. Karandashev and B.V. Kryzhanovsky,**

Center of Optical Neural Technologies,
Scientific Research Institute for System Analysis, Russian Academy of Sciences,
Nakhimovsky ave, 36-1, Moscow, 117218, Russia
malsagov@niisi.ras.ru, karandashev@niisi.ras.ru, kryzhanov@gmail.com

**Abstract.** Numerical methods are used to examine the thermodynamic characteristics of the two-dimensional Ising model as a function of the number of spins $N$. Onsager's solution is generalized to a finite-size lattice, and experimentally validated analytical expressions for the free energy and its derivatives are computed. The heat capacity at the critical point is shown to grow logarithmically with $N$. Due to the finite extent of the system the critical temperature can only be determined to some accuracy.

**Keywords:** partition function, free energy, critical point, heat capacity.

## 1. Introduction

The computation of the partition function is an important problem in statistical physics. The solution of the problem for a finite system will allow a noticeable advance in the methods of deep learning and image processing. Unfortunately, the exact solutions are found only for few models described in classical monographs [1, 2]. The advanced methods of statistical physics are used to investigate the associative memory properties [3–6] and develop the learning methods of finite neural networks [7–9].

A significant progress has been made in the development of numerical algorithms allowing successful investigation of critical characteristics [10-13] and energy spectra of spin systems [14]. The Monte Carlo approach [15, 16], which permits rough estimations, is mostly used in this kind of algorithms. However, algorithms [17-20] that make it possible to exactly calculate the free energy of a finite planar spin lattice have been developed. To best approximate the physical results corresponding to $N \to \infty$, researchers try to make calculations with as large number of spins $N$ as possible. However, the power of numerical computations is limited and the question of whether the dimensionality of the problem is large enough keeps open.

The aim of the paper is to study the relationship between system parameters and dimensionality $N$ of the problem and find analytical expressions suitable for finite $N$. The results given below make it possible to understand how large the dimensionality of the problem should be for the simulation results to give a satisfactory description of properties of real models. Besides, the analytical expressions suggest their use in deep training of finite neural networks and further development of image processing algorithms.

## 2. Basic expressions

Our interest is the free energy of the system:

$$f = -\ln Z / N, \tag{1}$$

where partition function $Z = \sum_S e^{-N\beta E(S)}$ is the sum over all possible spin configurations, $E = -\sum J_{ij} s_i s_j / 2N$ is the energy of the system, $s_i = \pm 1$, and $\beta$ is the inverse temperature. The knowledge of the free energy allows us to compute the major parameters of the system such as internal energy $U = \overline{E}$, energy variance $\sigma^2 = \overline{E^2} - \overline{E}^2$ and heat capacity $C = \beta^2 \sigma^2$:

$$U = \frac{\partial f}{\partial \beta}, \quad \sigma^2 = -\frac{\partial^2 f}{\partial \beta^2}, \quad C = -\beta^2 \frac{\partial^2 f}{\partial \beta^2} \tag{2}$$

In the experiment we use a planar model where spins are positioned in a square lattice and only four nearest neighbors interact $J_{ij} = J$. Onsager's solution [21] found for $N \to \infty$ for this sort of system with periodical boundary conditions has the form:

$$f(\beta) = -\frac{\ln 2}{2} - \ln(\cosh 2\beta J) - \frac{1}{2\pi} \int_0^\pi \ln\left(1 + \sqrt{1 - k^2 \cos^2 \theta}\right) d\theta, \tag{3}$$

where $k = 2\sinh 2\beta J / \cosh^2 2\beta J$. The solution describes the logarithmic divergence of heat capacity when $\beta \to \beta_{ONS}$, where the critical temperature is determined from condition $k = 1$ as:

$$\beta_{ONS} = \frac{1}{2}\ln(1 + \sqrt{2}). \tag{4}$$

### 3. The experimental results

We make an intensive use of the Kasteleyn-Fisher algorithm [17, 18] here to compute the free energy of the 2D square spin system. The algorithm gives exact results because the finding of the partition function is reduced to computation of the determinant of a matrix generated in accordance with the model under consideration. The algorithm permits us to exactly calculate the free energy of a spin system for an arbitrary planar graph with arbitrary links in a polynomial time. More information about the algorithm can be found in [19]. In the paper we use the realization [20] of the algorithm that can give the same results in a shorter time. Using this algorithm, we were able to examine the behavior of free energy and its derivatives (internal energy $U$ and heat capacity $C$) for a few lattices of different dimensions $N = L \times L$. The length of the lattice varied from $L = 25$ to $L = 10^3$. Let us point out that the algorithm we use is only applicable to planar lattices. It means that we considered only lattices with free boundary conditions because lattices with periodic boundary conditions do not belong to a planar graph.

Correspondingly, the basic state energy is

$$E_0 = -2\left(1 - \frac{1}{\sqrt{N}}\right). \tag{5}$$

In the experiment we computed free energy $f = f(\beta)$ and its derivatives. As expected, the peak of curve $C = C(\beta)$ is shifted to the right from the peak of curve $\sigma = \sigma(\beta)$. The position of the heat capacity peak is used determine critical temperature $\beta_c$ and critical values $f_c = f(\beta_c)$, $U_c = U(\beta_c)$, $\sigma_c = \sigma(\beta_c)$ and $C_c = C(\beta_c)$. The position of the energy variance peak is used to find the second critical point $\beta_c^*$ and corresponding critical values $f_c^* = f(\beta_c^*)$, $U_c^* = U(\beta_c^*)$, $\sigma_c^* = \sigma(\beta_c^*)$ and $C_c^* = C(\beta_c^*)$. All of these values are given in Table 1.

The results of the experiment and data analysis are presented graphically in Figures 1-4. As seen from Fig. 1, the experimentally found values of free energy and internal energy approach Onsanger's solution with growing dimensionality. The figure gives the curves only for small lengths $L = 25, 50, 100$. When $L > 100$, the curves practically repeat Onsanger's solution and are not shown in the Figure for this reason. According to (5), the asymptotical behavior of free energy for large $\beta$ is described as

$$f \approx -2\beta\left(1 - \frac{1}{\sqrt{N}}\right). \tag{6}$$

It is the presence of term $\sim 1/\sqrt{N}$ that causes the curves representing small linear dimensions not to follow Onsager's solution.

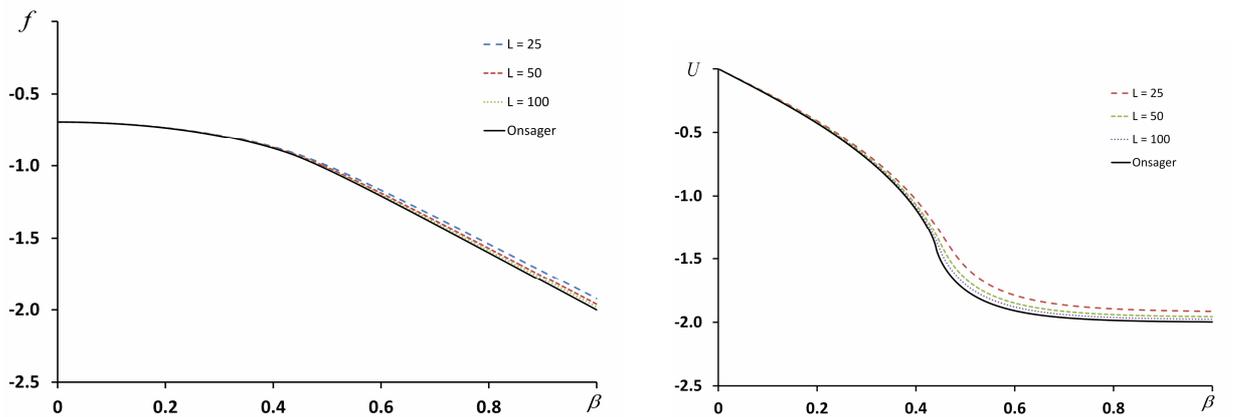

**Fig. 1.** Free energy $f = f(\beta)$ (the left plot) and internal energy $U = U(\beta)$ (the right plot) for small-dimension lattices and asymptotic Onsager's solution ($L \to \infty$).

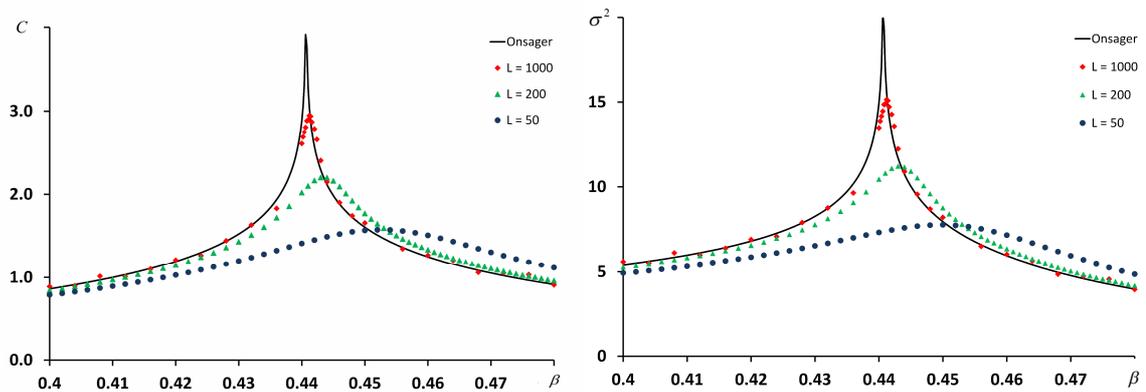

**Fig. 2**. Dependences of heat capacity $C$ (the left plot) and variance $\sigma^2$ (the right plot) from $\beta$ for some lattice dimensions from $N = 50 \times 50$ to $N = 10^3 \times 10^3$ (dots). Osanger's solution is drawn in solid.

For the finite-extent lattice the difference between Osanger's solution and $f = f(\beta)$ is best seen in the plot of the second derivative (heat capacity $C$). According to Osanger's solution the heat capacity exhibits logarithmic divergence when $\beta \to \beta_{ONS}$. In the case of finite lattice it does not happen; however, we can observe the peak in the heat capacity curve $C = C(\beta)$, the peak becoming sharper with the growing lattice dimension (see fig. 2). A closer examination shows that the peak height increases logarithmically with the lattice dimension and the peak itself is slightly shifted to the right from $\beta_{ONS}$, the distance between the two points shortening as the lattice dimension grows.

**Table 1.** Critical values at the peaks of heat capacity and energy variance.

| $L$ | $\beta_c / \beta_c^*$ | $f_c / f_c^*$ | $U_c / U_c^*$ | $\sigma_c / \sigma_c^*$ | $C_c / C_c^*$ |
|---|---|---|---|---|---|
| 25 | 0.4642 / 0.4556 | 0.9467 / 0.9351 | 1.3808 / 1.3288 | 2.4444 / 2.4678 | 1.2875 / 1.2642 |
| 50 | 0.4522 / 0.4494 | 0.9382 / 0.9344 | 1.3985 / 1.3768 | 2.7762 / 2.7849 | 1.5760 / 1.5664 |
| 100 | 0.4462 / 0.4454 | 0.9337 / 0.9326 | 1.4054 / 1.3978 | 3.0767 / 3.0782 | 1.8846 / 1.8797 |
| 200 | 0.4436 / 0.4432 | 0.9320 / 0.9314 | 1.4120 / 1.4075 | 3.3491 / 3.3502 | 2.2072 / 2.2046 |
| 300 | 0.4428 / 0.4422 | 0.9315 / 0.9306 | 1.4152 / 1.4078 | 3.4990 / 3.5001 | 2.4005 / 2.3955 |
| 400 | 0.4422 / 0.4418 | 0.9309 / 0.9304 | 1.4143 / 1.4091 | 3.6050 / 3.6052 | 2.5413 / 2.5369 |
| 500 | 0.4418 / 0.4418 | 0.9305 / 0.9305 | 1.4131 / 1.4131 | 3.6832 / 3.6832 | 2.6479 / 2.6479 |
| 600 | 0.4414 / 0.4414 | 0.9301 / 0.9301 | 1.4104 / 1.4104 | 3.7525 / 3.7525 | 2.7435 / 2.7435 |
| 700 | 0.4414 / 0.4414 | 0.9302 / 0.9302 | 1.4124 / 1.4124 | 3.8141 / 3.8141 | 2.8344 / 2.8344 |
| 800 | 0.4414 / 0.4414 | 0.9302 / 0.9303 | 1.4139 / 1.4139 | 3.8544 / 3.8544 | 2.8945 / 2.8945 |
| 900 | 0.4414 / 0.4414 | 0.9303 / 0.9303 | 1.4152 / 1.4152 | 3.8702 / 3.8702 | 2.9184 / 2.9184 |
| 1000 | 0.4412 / 0.4412 | 0.9301 / 0.9301 | 1.4132 / 1.4132 | 3.8914 / 3.8914 | 2.9477 / 2.9477 |

The examination of the data of table 1 shows that the position of the peak (critical value $\beta_c$) and the dimension dependencies of the critical values of free energy and heat capacity can be approximated well by the following expressions:

$$\beta_c = \beta_{ONS}\left(1 + \frac{5}{4\sqrt{N}}\right)$$

$$U_c = -\sqrt{2} \cdot \left(1 - \frac{1}{2\sqrt{N}}\right) \qquad (7)$$

$$C_c = \frac{4\beta_c^2}{\pi}\left(\ln N - 1.7808\right)$$

The approximation (7) of dependency $\beta_c = \beta_c(N)$ gives a rather small relative error: the largest error of ~0.3% is at $L = 25$, the error decreases rapidly with growing $L$ (to 0.01% at $L = 10^3$). The relative error of the approximation of $U_c$ is less than 0.4% and $C_c$ less than 0.8%. Fig. 3 shows how close expressions (7) follow experimental data.

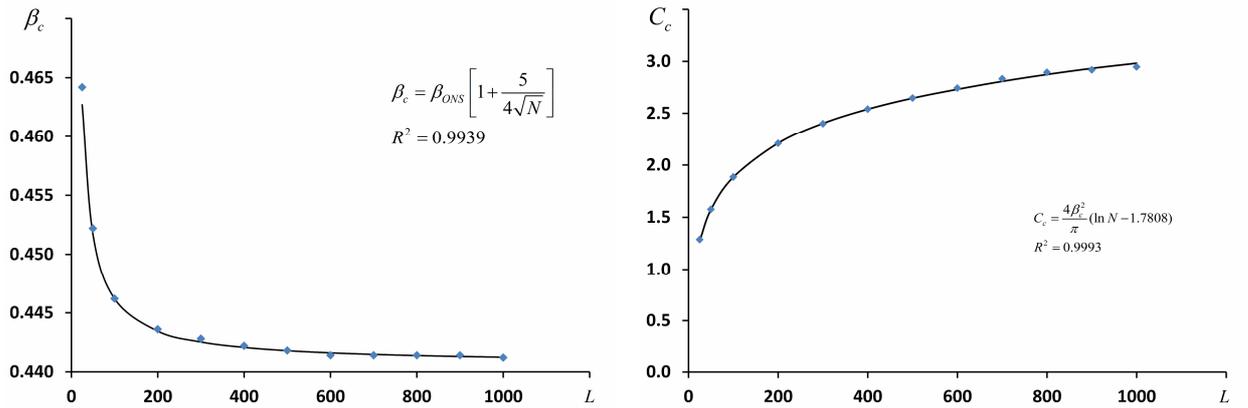

**Fig. 3**. Critical temperature $\beta_c$ (the left plot) and heat capacity $C_c$ (the right plot) as functions of dimension $L$: the dots represent experimental data, the solid lines approximation formulae (7).

The position of the energy variance peak (critical value $\beta_c^*$) and the corresponding values of free energy and heat capacity are well approximated (see table 1) by the expressions:

$$\beta_c^* = \beta_{ONS}\left(1 + \frac{1}{\sqrt{N}}\right)$$

$$U_c^* = -\sqrt{2} \cdot \left(1 - \frac{1}{\sqrt{N}}\right) \qquad (8)$$

$$C_c^* = 1.197\, \beta_{ONS}^2 \cdot \left(\ln N - 1\right)$$

These formulae give good agreement with experimental data: $\beta_c^*$, $U_c^*$ and $C_c^*$ have the greatest relative error 0.6%, 2.1% and 1.2% correspondingly at $L = 25$. The relative errors fall rapidly with $L$ and at $L = 10^3$ become 0.02%, 0.03% and 0.08%.

### 4. Generalization of Onsanger's solution

The analysis shows that it is possible to get the analytical expressions that can describe experimental data and the above approximation formulae quite well. It is sufficient to make substitutions $2\beta J \to z$ and $k \to \kappa$ in (3), where

$$z = \frac{2\beta J}{1+\Delta}, \quad \kappa = \frac{2\sinh z}{(1+\delta)\cosh^2 z}. \tag{9}$$

Then for free energy, internal energy and heat capacity we get

$$f(\beta) = -\frac{\ln 2}{2} - \ln(\cosh z) - \frac{1}{2\pi}\int_0^\pi \ln\left(1 + \sqrt{1 - \kappa^2 \cos^2\theta}\right)d\theta$$

$$U = -\frac{1}{1+\Delta}\left\{2\tanh z + \frac{\sinh^2 z - 1}{\sinh z \cdot \cosh z}\left[\frac{2}{\pi}\mathbf{K}_1 - 1\right]\right\} \tag{10}$$

$$C = \frac{z^2}{\pi \tanh^2 z} \cdot \left\{a_1(\mathbf{K}_1 - \mathbf{K}_2) - (1 - \tanh^2 z)\left[\frac{\pi}{2} + (2a_2\tanh^2 z - 1)\mathbf{K}_1\right]\right\}$$

where $\mathbf{K}_1 = \mathbf{K}_1(\kappa)$ and $\mathbf{K}_2 = \mathbf{K}_2(\kappa)$ are full elliptical integrals of first and second type correspondingly and

$$a_1 = p(1+\delta)^2, \quad a_2 = 2p - 1, \quad p = \frac{(1-\sinh^2 z)^2}{(1+\delta)^2 \cosh^4 z - 4\sinh^2 z}. \tag{11}$$

As could be expected, when $N \to \infty$, formulae (11) give $p \to 1$, $a_{1,2} \to 1$ and expressions (10) turn into well-known ones [1, 2]. If we compare resulting expressions (10) with experimental data, we can see that the best agreement occurs when the adjustment parameters take the form:

$$\Delta = \frac{5}{4\sqrt{N}}, \quad \delta = \frac{\pi^2}{N} \tag{12}$$

Expressions (10) give good approximation of experimental results even if the lattice dimension is small. By way of illustration figure 4 gives the curves of energy variance and heat capacity for a $N = 25 \times 25$ lattice. It is seen that there is good agreement between the theory and experiment. This agreement becomes better with the growing lattice dimension.

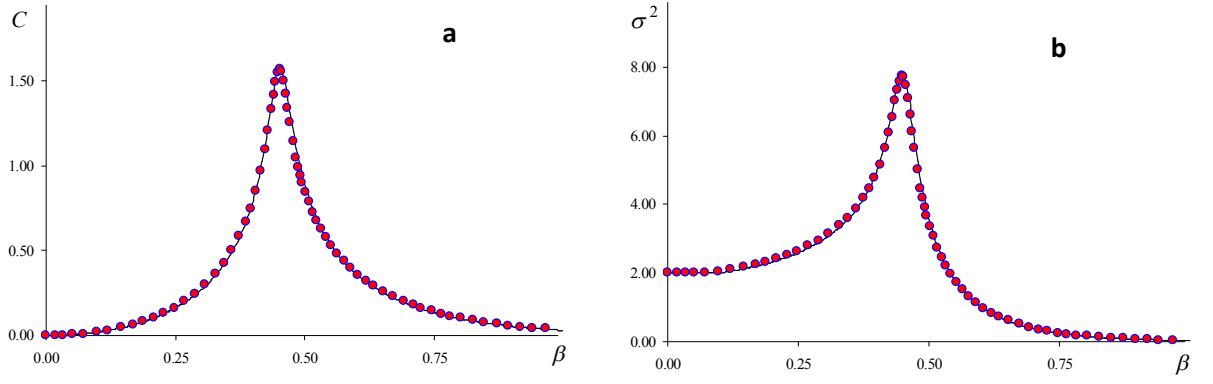

**Fig. 4.** Dependencies of heat capacity (the left plot) and variance (the right plot) on $\beta$ for a $N = 25 \times 25$ lattice: solid-line curves are produced by formulae (10), dots are experimental data.

The examination of expressions (10) shows that the introduction of the correction for a finite lattice dimension does not change behavior of free and internal energy much. On the other hand, in the heat capacity formula the logarithmic divergence at the critical point disappears. Indeed, the examination shows that the maximum of heat capacity occurs at $\sinh z = 1$, which corresponds to the critical temperature from $\beta_c = \beta_{ONS}(1 + \Delta)$. Borrowing $\Delta$ from (12), we find that the expression agrees with empirically defined expression (7) fully.

Expanding function $C(\beta)$ in a series about critical point $\beta_c$ and omitting the terms that are polynomial in $(\beta - \beta_c)$, we

$$C(\beta) \approx \frac{4\beta_c^2 J^2}{\pi} \left\{ 3\ln 2 - \frac{\pi}{2} - \ln\left[ 4J^2(\beta - \beta_c)^2 + \frac{\pi^2}{N} \right] \right\} . \qquad (13)$$

This expression yields the following expression for the critical heat capacity:

$$C_c = \frac{4\beta_c^2 J^2}{\pi} \left( \ln N + 3\ln 2 - 2\ln \pi - \frac{\pi}{2} \right), \qquad (14)$$

which corresponds to (7) because $2\ln \pi + \pi/2 - 3\ln 2 \approx 1.7808$.

The availability of expressions (10) allow us to examine other characteristics, e.g. the correlation and spontaneous magnetization. Let us first consider $N$-dependence of correlation length $\xi$. With $N \to \infty$ it is defined by the well-known expression [1], which we represent as $\xi = -1/2\ln \eta$, where $\eta = k/(1 + \sqrt{1 - k^2})$. To pass to the case of a finite lattice, let us make substitution $k \to \kappa$ as in (9). Then for the correlation length we get

$$\bar{\xi} = -\frac{1}{2\ln \bar{\eta}}, \qquad \bar{\eta} = \frac{\kappa}{1 + \sqrt{1 - \kappa^2}} . \qquad (15)$$

At critical point $\beta = \beta_c$ quantity $\kappa = \kappa(z)$ reaches its maximum $\kappa_{max} = 1/(1+\delta)$, and the correlation length gets greatest value $\xi_{max} = L/2\pi\sqrt{2}$, which is an order of magnitude as less as linear dimension $L$.

Similarly, let us make substitution $k \to \kappa$ in the expression [22] obtained by Yang. Then for the spontaneous magnetization we get

$$M_0 = \left(1 - \overline{\eta}^2 / \overline{\eta}_0^2\right)^{1/8}, \qquad \overline{\eta}_0 \approx 1/(1+\sqrt{2\delta}) \qquad (16)$$

($M_0 = 0$ when $\beta < \beta_c$). We should point out that interpretation of expression (16) with respect to a finite-extent system differs from the limiting case of $N \to \infty$ radically. As mentioned in [1], the mean magnetization of a finite-extent system in the absence of external field is zero because any configuration with $s_i = +1$ has its equiprobable counterpart with $s_i = -1$. Correspondingly, expression (16) can be used to describe the following fact: the magnetization measured in the experiment at different moments can take any value from $M_0$ and $-M_0$.

## 5. Discussion and conclusions

Using experimental data, we have formed simple evaluating expressions (7)-(8) for critical points. The expressions agree with experimental data quite well. Of course, we might have refined the adjustment coefficients in these expressions and derived evaluating expressions to better accuracy; yet it was not our goal. The primary aim of the research was to understand how the behavior of critical parameters would vary with $N$.

By introducing two adjusting parameters (12), we have built analytical expressions (10) extending Onsager's solution to a finite-dimension lattice. These formulae describe the behavior of a spin system very accurately even with small lattice dimensions ($N \sim 25 \times 25$). Given $N \geq 50 \times 50$, the disagreement with experimental data becomes less than the error of the experiment.

The research allows us to draw the following conclusions.

First, a simulation experiment usually makes it possible to correctly determine the behavioral features of a spin system even if $N$ is relatively small. The increasing of $N$ only helps to define critical parameters more precisely. However, this correction is not so important. Indeed, according to (7), the accuracy of determination of critical parameters $\beta_c$ and $U_c$ is dependent on $\sim L^{-1}$. It is hard to speak about $C_c$ because $C_c \sim \ln N$ and the increase of dimensionality by an order of magnitude from $L = 10^3$ to $L = 10^4$ results in an uninformative ~30% change of $C_c$.

Second, at the critical point finite-extent systems do not have the logarithmic divergence of heat capacity predicted by Onsanger. The same is true for the energy variance. And this should have been expected from most general speculations. Instead, we observe the logarithmic growth of the critical value of heat capacity like $C_c \sim \ln N$. It would seem that with $N \to \infty$ we go to Onsager's limit: $C \to \infty$ when $\beta \to \beta_c$. However, it is difficult to be done in practice: even if we make the dimensionality as great as Avogadro's number ($N \sim 10^{23}$), we get $C_c$ only four times as large as that for $N = 10^6$ (the case under consideration). Moreover, the dependence of heat capacity on $N$ means the observable (even at $N \sim 10^{23}$) violation of the additivity concept of a classical system: a twofold increase of the system dimensionality leads to a 1.3% growth of $C_c$.

Third, the experiment has shown that the peaks of curves $\sigma = \sigma(\beta)$ and $C = C(\beta)$ do not coincide: the heat capacity peak occurs at greater values of $\beta$. This is an expected result because the heat capacity and energy variance are related as $C(\beta) = \beta^2 \sigma^2(\beta)$. However, the question arises as to which of the peaks should be used to determine the critical temperature. Really, the first peak corresponds to the maximum of energy variance, the second to the maximum of correlation length. The both are indications of a phase transition. Conventionally, we use the heat capacity peak to determine the critical point. Given large system dimensionality, the approach is fully justified: when $N \geq 400 \times 400$, the separation between the peaks is almost unobservable. However, with smaller $N$ the gap between the peaks is quite noticeable. When interpreting experiments involving small-dimensionality lattices, we should rather say that the phase transition is spread over the temperature range from $\beta_c^*$ to $\beta_c$. It means that the numerical experiment allows us to determine the critical temperature to within the length of this range, i.e. the absolute error is to be of the order of $\pm \beta_{ONS} / 4L$.

The relation between the critical parameters and lattice dimensionality is considered using a two-dimensional Ising model. Yet we think that the key conclusions will be also true for other models.

The research is partially supported by the Russian Foundation for Basic Research (grants 16-31-00047 and 15-07-04861).